\documentclass[twocolumn,aps,prl,showpacs,superscriptaddress]{revtex4-2}

\usepackage{amsmath}
\usepackage{amsfonts}
\usepackage{amssymb}
\usepackage{amsmath}
\usepackage{amsthm}
\usepackage{mathrsfs}
\usepackage{graphicx}
\usepackage{color}
\usepackage{amsfonts}
\usepackage{wrapfig}
\usepackage{cases}
\usepackage{framed}
\usepackage{float}
\usepackage[colorlinks,citecolor=blue]{hyperref}
\usepackage{multirow}

\usepackage{bigints}
\usepackage[mathabx]{varint}

\begin{document}

\title{Quantum-Enhanced Sensing of Excited-State Dynamics with Correlated Photons}



\author{Jiahao Joel Fan}
\affiliation{Department of Physics, City University of Hong Kong, Kowloon, Hong Kong SAR}

\author{Feihong Liu}
\affiliation{Department of Materials Science \& Engineering, City University of Hong Kong, Kowloon, Hong Kong SAR}

\author{Dangyuan Lei}
\affiliation{Department of Physics, City University of Hong Kong, Kowloon, Hong Kong SAR}
\affiliation{Department of Materials Science \& Engineering, City University of Hong Kong, Kowloon, Hong Kong SAR}

\author{Shaul Mukamel}
\email{smukamel@uci.edu}
\affiliation{Department of Chemistry, University of California, Irvine, Irvine, CA 92697, United States}

\author{Zhedong Zhang}
\email{zzhan26@cityu.edu.hk}
\affiliation{Department of Physics, City University of Hong Kong, Kowloon, Hong Kong SAR}
\affiliation{City University of Hong Kong, Shenzhen Research Institute, Shenzhen, Guangdong 518057, China}

\date{\today}

\begin{abstract}
Squeezed photons, which provides a quantum-correlated light with reduced noise, have emerged as a powerful resource for sensing the structures of matter. Here, we study the pump-probe scheme using squeezed photons whose spectral correlation of amplitudes can be tailored. A microscopic theory reveals a highly time-energy-resolved signal that is not attainable by classical pump-probe scheme. 
An application to a monolayer semiconductor model reveals a real-time monitoring of valley excitons and their dynamics. 
We further show the intermediate squeezing regime--not strong squeezing--where time-resolved spectroscopy is favorable. Our work offers a new paradigm for studying nonequilibrium dynamics of matter, in photocatalysis and optoelectronics.
\end{abstract}

\maketitle

{\it Introduction.}---Squeezed photons are receiving great attention in studies of spectroscopy and sensing \cite{PhysRevLett.58.2539, PhysRevLett.68.3020, Polzik, PhysRevA.46.R6797, doi:10.1126/science.1097576,  Qu:13,PhysRevA.93.053802, npj, Junker:21,  Prajapati:21, Li:22,  PhysRevA.106.023115,Cutipa:22, PhysRevLett.130.133602}. The squeezed light, as a typical nonclassical state of photons, is notable for compressing intensity/noise fluctuations below the standard quantum limit, not attainable by classical light \cite{Yurke:87,PhysRevLett.75.3426}. This has led to rapid development of quantum technologies including the precision measurements, memory and information processing. Multi-mode squeezing can bring out remarkable quantum correlation, in addition to noise reduction overcoming the shot-noise limit \cite{PhysRevLett.55.2409, doi:10.1126/science.1104149, PhysRevApplied.15.044030}. The correlation may open up an extra dimension for controlling the spectral scale of the optical signals. Recent studies showed the ability to combine the quantum correlation with field brightness--the bright squeezing--which led to successful developments in various fields, e.g., gravitational-wave detection \cite{Goda:2008ie, Oelker:14, article, Mehmet_2019, article2}, optical interfermetery \cite{PhysRevLett.59.2153, PhysRevLett.59.278, PhysRevLett.88.231102,  PhysRevLett.95.211102, SCHNABEL20171}, magnetometers \cite{PhysRevLett.105.053601, PhysRevA.86.023803}, and electronic microscopy \cite{PhysRevX.4.011017, 10.1063/5.0009681, Casacio2021QuantumenhancedNM, Xu:22, Terrasson2024FastBI, photonics11060489}.

Earlier, quantum-light spectroscopy exploited the entangled photon pairs, yielding a multidimensional sensing of molecular excited states by means of quantum interference \cite{PhysRevLett.78.1679,guzman2010spatial, PhysRevLett.80.3483, PhysRevLett.93.023005,doi:10.1021/jp066767g,doi:10.1021/acs.jpca.7b06450, PhysRevA.76.043813,doi:10.1021/ja803268s,doi:10.1021/acs.jpca.8b06312, PhysRevLett.123.023601, doi:10.1063/5.0049338,doi:10.1021/jacs.1c02514,doi:10.1021/acs.accounts.1c00687,doi:10.1021/acsphotonics.2c00255,doi:10.1063/5.0128249,RevModPhys.88.045008, Zhang2022, Schlawin1, Fan2023EntangledPE}. This scheme however encounters a tremendous challenge of ultra-weak signal intensity, causing a frustration in signal detection \cite{fujihashi}. One way out of this difficulty is to use brighter light beams, which revokes the squeezed photons. Several developments had called for the enhanced coupling of molecular excitations to squeezed light field, demonstrating the quantum supremacy in nonlinear optical processes \cite{Yurke:87, Leonski:93, PhysRevLett.58.2539, PhysRevLett.56.1917, PhysRevA.61.033811, npj, Li:22, Qu:13,  PhysRevX.6.031004, PhysRevA.94.063825, PhysRevA.106.023115, Cutipa:22, Schlawin2017, Hardal:19, PhysRevA.106.023705,PhysRevLett.134.133604}. For instance, efficient two-photon absorption has been observed by applying squeezed photon beams onto biomarkers \cite{GSA_APL2020}. Recent experimental advancements have resulted in the quantum-correlated ultrashort pulses, enabling a quantum-correlated transient absorption spectroscopy \cite{doi:10.1126/sciadv.adt2187}. While research is ongoing to achieve the spectral measurement, its joint time-energy resolution is ultimately limited by the pulse parameters and their correlations. This may yield a lower bound of time-energy scale, which still remains elusive. These bottlenecks may constitute limits in studying the highly-excited dynamics including the nonadiabatic transitions in organic semiconductors and molecular aggregates, which place urgent call for new spectroscopic techniques.

In this article, we develop a pump-probe-fluorescence (PPF) spectroscopy using squeezed photons. This addresses the key question: can quantum correlation give the resolution beyond the bound of laser pulses, i.e., $\Delta E \Delta t < \hbar$? We find that achievable via quantum correlation between light amplitudes which balances off the squeezing. A microscopic theory is developed, whereby the two correlated photon fields serve as the pump and probe fields. The real-time sensing of excited-state dynamics is thus exemplified by a monolayer semiconductor model--the exciton states and their conversion are clearly resolved. The signal further suppresses ground-state bleaching (GSB) component from the quantum correlation. 
Finally, the signal intensity is significantly amplified by the squeezed coherent photons, surpassing the bottleneck of using entangled photon pairs.

\begin{figure}[t]
\centering
\includegraphics[width=0.48\textwidth,height=8cm]{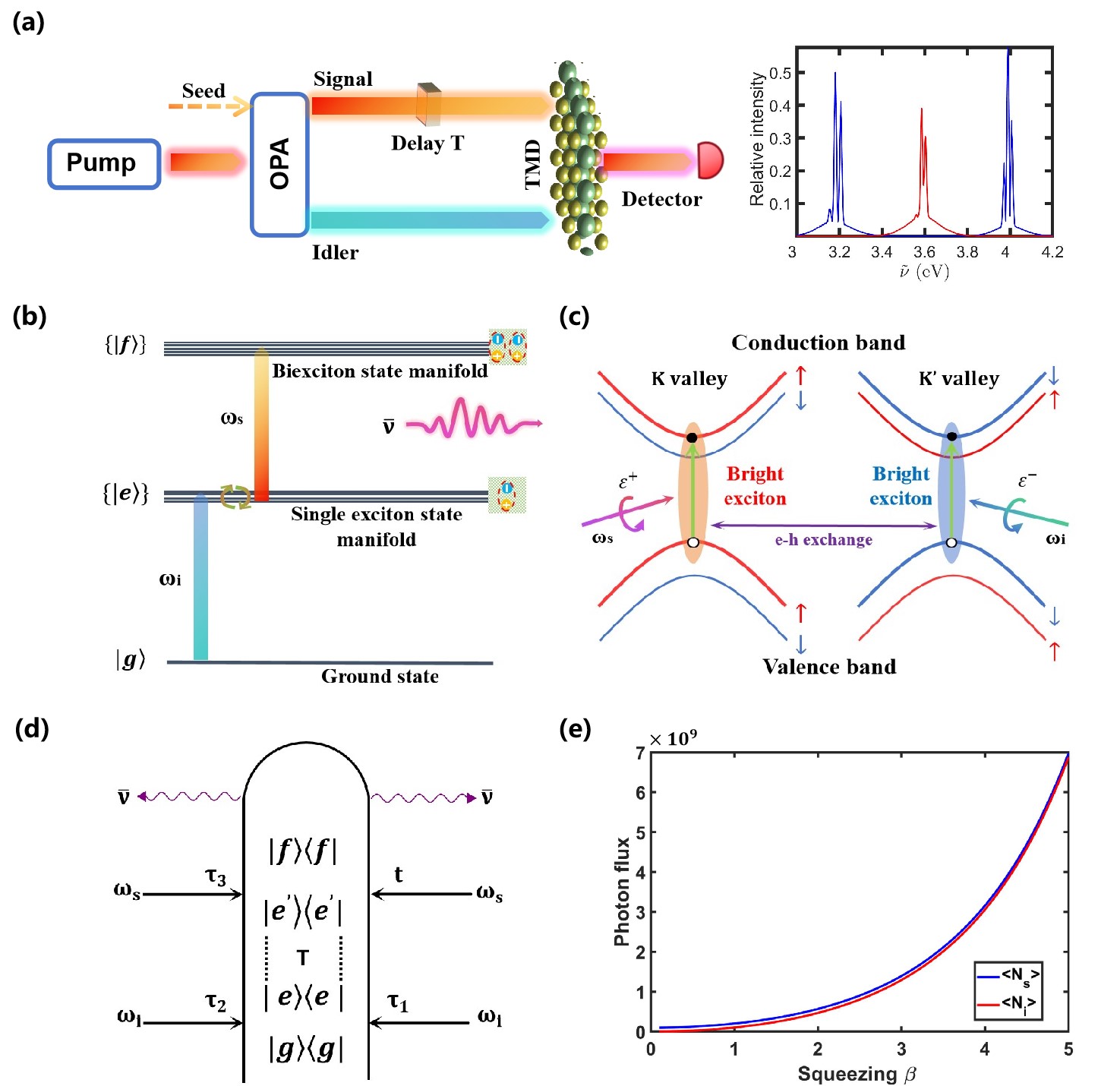}
\caption{Illustration of optical signal generated by materials interacting with correlated photons. (a) Pump-probe-fluorescence scheme with squeezed photons where the optical parametric amplifier (OPA) box is seeded by a supercontinuum white light in order to generate the squeezed coherent state; Small panel plots the uncertainties of measurements of the squeezed light. (b) Level scheme for the PPF with squeezed light. (c) K-K$'$ valleys of the monolayer semiconductors, e.g., WS$_2$. (d) Feynman's loop diagram for the pump-probe-fluorescence signal. (e) Photon fluxes in s and i arms, with $|\alpha|^2=10^4$.}
\label{Schematic}
\end{figure}

{\it General formalism for quantum-light PPF.}---Given an optical parametric amplifier (OPA) that generates the s and i beams, the intensity/phase of the two beams is squeezed, resulting in a spectral correlation between the two beams. In the pump-probe scheme, photons in s and i arms serve as the respective pump and probe fields interacting with the system [see Fig.\ref{Schematic}(a)]. To have a neat picture, we consider the dipolar coupling, i.e.,
\begin{equation}
  \begin{split}
    H_{{\rm int}}(t) = - {\bf V}^+(t) \cdot \big[{\bf E}_s(t) + {\bf E}_i(t) \big] - \text{h.c.}
  \end{split}
\label{Hint}
\end{equation}
where ${\bf V}^+(t)$ is the raising part of the dipole operator; ${\bf E}_a(t); a=s,i$ is the negative-frequency part of the electric field in each arm containing multiple frequency modes. The dynamics follows $|\Psi_{{\rm f},{\rm k}}(t)\rangle = \hat{{\cal T}} e^{-{\rm i}\int^t H_{{\rm int}}(t')\text{d}t'}|\Psi_{{\rm i},{\rm q}}(t_0)\rangle \otimes |\Phi\rangle$ where $\Psi_{{\rm i},{\rm q}}$ and $\Phi$ are the initial states of system and field, respectively. 

One has to collect the fluorescent photons from the ${\rm f}$ states, as depicted in Fig.\ref{Schematic}(b), yielding the energy-resolved signal $S(v) = \langle a_v^{\dagger}a_v\rangle$ where $a_v$ is the annihilation operator of the emitted photons at frequency $v$. The PPF is generated by a subsequent action of the s and i beams. 
Using Dyson's equation against $H_{{\rm int}}(t)$, as illustrated by the loop diagram in  Fig.\ref{Schematic}(d), one obtains \cite{Schlawin_JPB2013}
\begin{equation}
  \begin{split}
    S(v) \propto \bigintssss [\text{d}\tau] e^{{\rm i}v \bar{\tau}}  \chi(\tau_1 - t, \tau_3 - \tau_2) C(t,\tau_3,\tau_2,\tau_1)
  \end{split}
\label{signal}
\end{equation}
where $[\text{d}\tau] \equiv \prod_{n=1}^4 \text{d}\tau_n$ and $\bar{\tau} = \bar{\tau}_2 - \bar{\tau}_1$, $\bar{\tau}_a = \tau_a - T_s$; $T_s$, $T_i$ are the arrival times of the s and i photons. $\chi(\tau_1 - t, \tau_3 - \tau_2) = \langle \Psi_{{\rm i},{\rm q}}|\text{V}^- \text{U}(\tau_1 - t) \text{V}^- \text{V}^+ \text{U}(\tau_3 - \tau_2) \text{V}^+|\Psi_{{\rm i},{\rm q}} \rangle$ is the time-domain response function of the materials, where $|\Psi_{{\rm i},{\rm q}}\rangle$ is the ground-state wave function of the materials. ${\rm U}(t)$ is the operator propagating the excited-state dynamics once the material has been pumped.

The four-point Green's function of fields for the squeezed photons
\begin{equation}
  \begin{split}
    & C(t,\tau_3,\tau_2,\tau_1) = \\[0.15cm]
    & \langle \Phi|E_i^{\dagger}(\tau_1-T_i)E_s^{\dagger}(t-T_s)E_s(\tau_3-T_s)E_i(\tau_2-T_i)|\Phi\rangle
  \end{split}
\label{C4}
\end{equation}
with $|\Phi\rangle = |\alpha,z\rangle$ may manifest the quantum correlation, as a key for unusual spectro-temporal properties of the quantum-light PPF. 
$|\alpha,z\rangle = {\cal S}(z)|\alpha\rangle$ denotes the squeezed coherent state of photons, where $|\alpha\rangle$ is the coherent state from the seed pulse, and ${\cal S}(z) = e^{-\iint\text{d}\omega_s \text{d}\omega_i [z\Phi(\omega_s,\omega_i) a_{\omega_s}^{\dagger} a_{\omega_i}^{\dagger} - \text{h.c.}]}$
is the multi-mode squeezing operator; $z = \beta e^{{\rm i}\theta}$. $\Phi(\omega_s,\omega_i)$ is the photon wave function describing the down conversion process inside the OPA; $a_{\omega_s}^{\dagger}$ and $a_{\omega_i}^{\dagger}$ create the photons in s and i arms, respectively.

{\color{red}It is worth noting that the GSB component involves a spectral resonance, i.e, $\sim 2 \omega_{\rm eg}$ away from the one $\sim \omega_{\rm fg} -\omega_{\rm e'e}$ in Fig.\ref{Schematic}(d). $\omega_{\rm eg}$ ($\omega_{\rm fg}$) is the energy gap between ground and singly-excited (doubly-excited) states;  $\omega_{\rm e'e}$ is the energy splitting between singly-excited states. All details are given in SM. The GSB intensity can be controlled, namely, either being suppressed or enhanced}.

{\it Quantum correlation of squeezed photons.}---To show the unusual properties of the correlation, we calculate $C(t,\tau_3,\tau_2,\tau_1)$ using the Bogoliubov transform \cite{Dorfman_RMP2016}, i.e.,
\begin{equation}
  \begin{split}
    & A_k \rightarrow \sqrt{G_k} A_k - e^{{\rm i}\theta}\sqrt{G_k - 1} B_k^{\dagger} \\[0.2cm]
    & B_k^{\dagger} \rightarrow - e^{-{\rm i}\theta}\sqrt{G_k - 1} A_k + \sqrt{G_k} B_k^{\dagger}
  \end{split}
\label{Bogoliubov}
\end{equation}
where $G_k = \cosh^2 (\beta r_k)$ is the optical gain and $A_k = \int \text{d}\omega_s \psi_k^* (\omega_s) a_{\omega_s}$, $B_k = \int \text{d}\omega_i \phi_k^* (\omega_i) a_{\omega_i}$. $\Phi(\omega_s,\omega_i) = \sum_k r_k \psi_k(\omega_s) \phi_k(\omega_i)$ from the singular-value-decomposition theorem, with $r_k$ as the mode weight.

The calculation proceeds with assuming the s port of the OPA is seeded by a supercontinuum white light $|\alpha\rangle = |\alpha\rangle_s \otimes | 0\rangle_i$. 
Then one obtains from Eq.(\ref{C4})
\begin{equation}
  \begin{split}
    & C (t,\tau_3,\tau_2,\tau_1) \propto Z(\alpha) \times \\[0.15cm]
    & \quad \Big[ H(\tau_2-T_s,\tau_3-T_i) H^*(\tau_1-T_s,t-T_i) + \cdots \Big]
  \end{split}
\label{ct}
\end{equation}
with the amplification factor
\begin{equation}
    Z(\alpha) = (1+|\alpha|^2)^2
\label{F}
\end{equation}
as a result of the optical gain for seed fields. $H(t_1,t_2)$ is the dual Fourier transform of $h(\omega_1,\omega_2)$, 
revealing the temporal correlation of photon amplitudes not attainable by classical light; $h(\omega_1,\omega_2) = 0.5\sum_k \psi_k(\omega_1) \phi_k(\omega_2) \sinh(2\beta r_k)$ reveals the spectral correlation of photon amplitudes. The nonresonant background (denoted by $\cdots$) contains two terms, arising from the quantum fluctuations of squeezed photons, that are smooth functions for an appropriate $\beta$.

\begin{figure}
\centering
\includegraphics[width=0.482\textwidth,height=6cm]{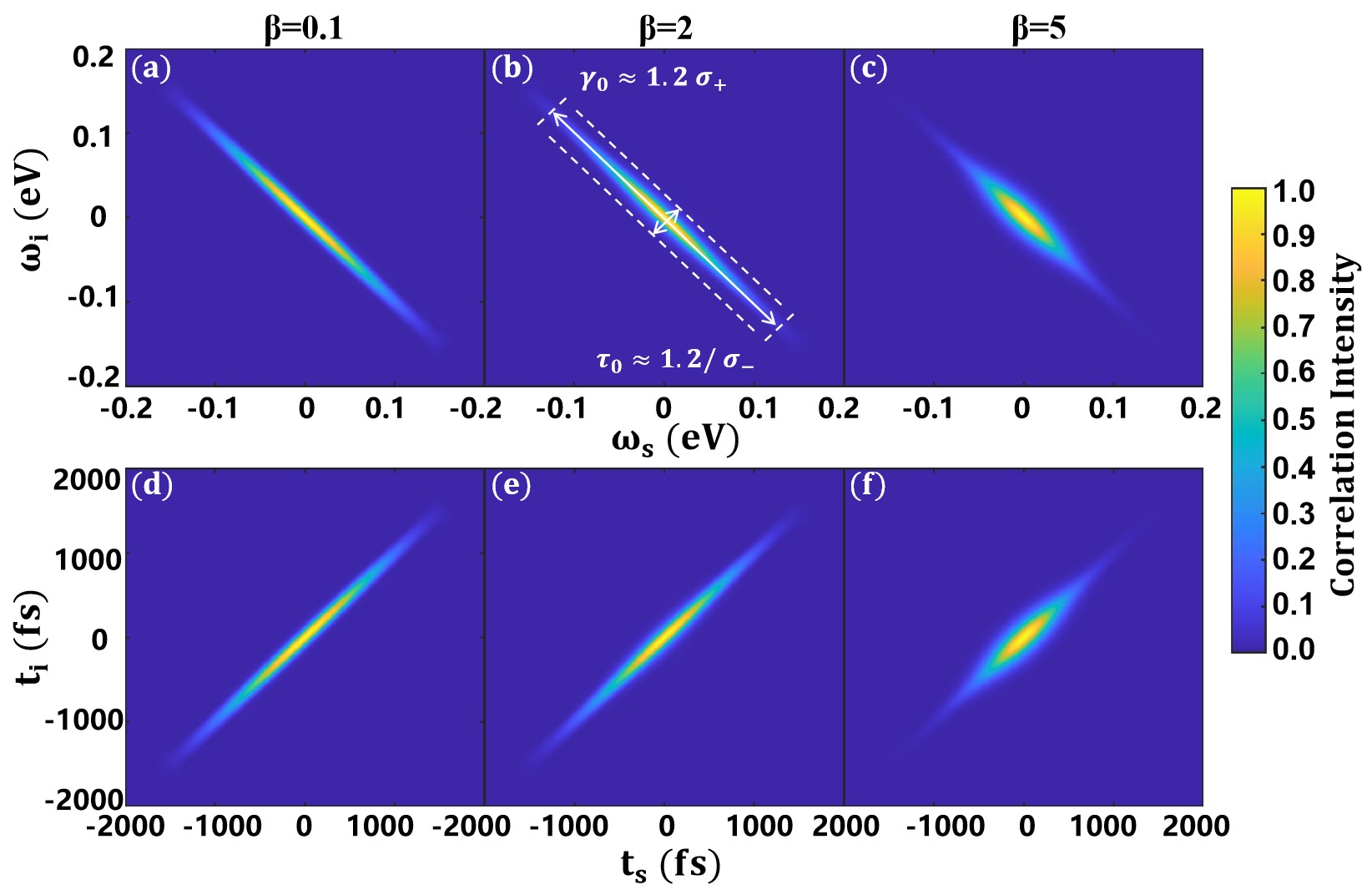}
\caption{The amplitude correlation of squeezed photons, based on Eq.(\ref{ct}). (Top) Spectral correlation $|h(\omega_s,\omega_i)|$ versus squeezing parameter $\beta$. (Bottom) Temporal correlation $|H(t_s,t_i)|$ versus $\beta$. In (b,e), $\tau_0 \gamma_0 = 0.09$. Parameters are $\sigma_+ = 5$meV, $\sigma_- = 82.7$meV, 
$\omega_{-}=0 $, where $\sigma_{\pm}$ and $ \omega_-$ are included in the photon wave function $\Phi(\omega_s,\omega_i)$.}
\label{hH}
\end{figure}

The cigar shape of $H(t_1,t_2)$,  $h(\omega_1,\omega_2)$ reflects the temporal and spectral correlations of photons, shown in Fig.\ref{hH}. This defines two parameters $\tau_0^{-1}, \gamma_0$ as measures of the anti-diagonal broadening and diagonal broadening of $h(\omega_1,\omega_2)$, which are controllable independently through the pump, seed and phase matching in the OPA. Likewise, for $H(t_1,t_2)$, the two broadenings are measured by $\gamma_0^{-1}$, $\tau_0$. Fig.\ref{hH}(b,e) show $\tau_0 \gamma_0 = 0.09$ for $\beta=2$.

Fig.\ref{hH} shows from the upper row that the photons in s and i arms are spectrally anti-correlated at weak ($\beta=0.1$) and intermediate ($\beta=2$) squeezing, corresponding to the squeezing of 0.2dB and 7dB, respectively. The anti-correlation becomes weaker at a strong squeezing $\beta=5$ for a squeezing $>10$dB. The spectroscopy is in favor of the intermediate squeezing regime, which may interrogate the time-energy scales and the signal intensity. The latter is evident by calculating the photon flux in each arm, i.e., $N_s = \sum_k [\sqrt{Z(\alpha)} G_k - 1]$, $N_i = \sum_k\sqrt{Z(\alpha)} (G_k - 1)$. Obviously $N_s - N_i = \sum_k |\alpha_k|^2$ gives the photon flux in the seed.

Fig.\ref{hH}(c,f) indicate that the pump-probe (two-photon absorption) scheme using squeezed light is indistinctive from the one using lasers, at a strong squeezing. This is because the temporal/spectral correlation between photon-wave amplitude is significantly eroded.

{\it PPF with squeezed photons.}---On substituting Eq.(\ref{ct}) into Eq.(\ref{signal}), one obtains the PPF signal
\begin{widetext}
\begin{equation}
  S(v,T) \propto Z(\alpha) \sum_{{\rm f},{\rm k}} \Bigg|\int_{\; 0}^{\infty} \text{d}t \langle \Psi_{{\rm f},{\rm k}}|{\rm V}^+ \text{U}(t) {\rm V}^+|\Psi_{{\rm i},{\rm q}}\rangle 
     e^{{\rm i}(\omega_{{\rm f},{\rm k}} + \omega_{{\rm i},{\rm q}} + \omega_-)\frac{t - T}{2}} {\cal M} \left(\omega_{{\rm f},{\rm k}} - \omega_{{\rm i},{\rm q}} - \bar{v}, \frac{T - t}{2\tau_0}\right) \Bigg|^2
\label{SPPF}
\end{equation}
\end{widetext}
with $\bar{v} \equiv v + \omega_+$ as the photon frequency in the fluorescence; $|\Psi_{{\rm f},{\rm k}} \rangle$ denotes the highly-excited states of the materials. The time delay $T = T_s - T_i$ is controllable via optical pathway and is crucial for reading out the real-time dynamics of materials. The 2D function $\left|{\cal H}(t_1,t_2/\tau_0) \right| = (2\pi)^2 \tau_0\left|H(t_1+t_2,t_1-t_2) \right|$ 
is a counter-clockwise $45^{\circ}$ rotation of the $H$ function. 
\begin{equation}
  {\cal M}\left(\omega, \frac{\tau}{\tau_0}\right) = \bigintssss_{-\infty}^{\infty} {\rm d}t\ e^{{\rm i}\omega t} {\cal H} \left(t,\frac{\tau}{\tau_0}\right)
\end{equation}
can thus be narrow in both time and energy domains. In Eq.(\ref{SPPF}) the nonresonant terms have been dropped, as they are smooth functions shown in SM.

Eq.(\ref{SPPF}) is an important result such that it reveals a simultaneously high temporal and spectral scales, i.e., $\delta\omega \delta T<1$ which can readily monitor the nonequilibrium dynamics. ${\cal M}(\omega,\tau/\tau_0)$ arising from the field correlation of the squeezed light ensures the energy conservation; upon the detection of fluorescent photons, the photon frequency and thus the states are fixed. Moreover, the squeezed light offers a time gate $T - \tau_0/2 \leqslant t \leqslant T + \tau_0/2$ for interrogating the excited-state dynamics. The gate length $\tau_0$ and the spectral width $\gamma_0$ of ${\cal M}(\omega,\tau/\tau_0)$ are subject to the respective anti-diagonal broadening and diagonal broadening of the field correlation, as depicted in Fig.\ref{hH}. These two parameters can therefore be unconjugated, yielding $\tau_0 \gamma_0 < 1$. To see this closely, Eq.(\ref{SPPF}) reduces to
\begin{equation}
 \begin{split}
   & S(v,T) \propto Z(\alpha) \times \\[0.15cm] & \sum_{{\rm f},{\rm k}} \Big|\langle\Psi_{{\rm f},{\rm k}}|\text{V}^+ \text{U}(T) \text{V}^+|\Psi_{{\rm i},{\rm q}}\rangle\ \Omega(\omega_{{\rm f},{\rm k}} - \omega_{{\rm i},{\rm q}} - \bar{v}) \Big|^2
 \end{split}
\label{SPPFn}
\end{equation}
under the {\it impulsive approximation}, i.e., $\tau_0\rightarrow 0$,  which holds for the exciton relaxation \& phonon dephasing much slower than $\tau_0$. $\Omega(\omega)$ is the line-shape function.

{\it Signal enhancement of sensing.}---The factor $Z(\alpha) \approx |\alpha|^4$ for $|\alpha|\gg 1$, namely, $|\alpha|^2 \sim 10^3 - 10^4$ available for the average intensity of the seed pulse in experiments. For the squeezed vacuum $|z\rangle = {\cal S}(z)|0\rangle$, one has $Z(0) = 1$. Therefore, the signal enhancement by using squeezed photons for the PPF is estimated as
\begin{equation}
  \frac{Z(\alpha)}{Z(0)} \approx |\alpha|^4 \sim 10^6 - 10^8.
\label{SE}
\end{equation}
Eq.(\ref{SE}) manifests a promising experimental feasibility of the quantum advantage, greater than the entangled photon pairs and the squeezed vacuum.

\begin{figure}
\centering
\includegraphics[width=0.48\textwidth,height=6.4cm]{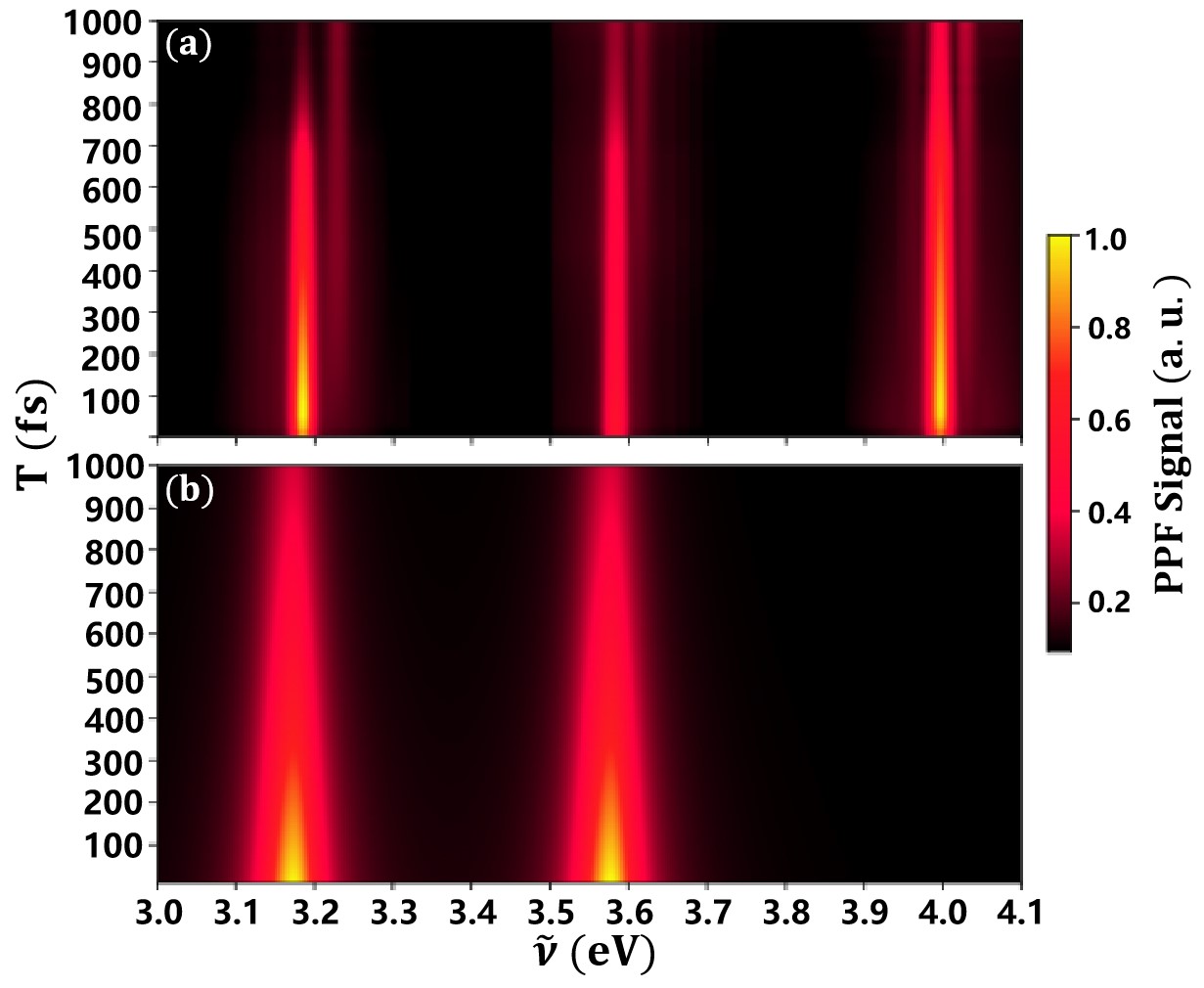}
\caption{(a) The PPF signal using squeezed coherent light from Eq.(\ref{SPPF}). The parameters are $\sigma_+ = 5$meV, $\sigma_-= 82.7$meV, $\omega_- = 0$ ($\omega_- = 0.42$eV) for side (medium) lines. (b) The PPF signal using laser pulses, where the bandwidth of pump and probe pulses is $\sigma_{{\rm p}},\ \sigma_{{\rm pr}} = 70$meV, corresponding to a temporal duration $60$fs. The PPF frequency $\tilde{\nu}=\omega_{{\rm p}}+\omega_{{\rm pr}}$ is obtained with pump pulse has a central frequency at $\omega_{{\rm p}} = 1.6$eV.}
\label{QPPF}
\end{figure}

{\it Exciton dynamics in monolayer model.}---We simulate the PPF in Eq.(\ref{SPPF}) for a monolayer semiconductor possessing a band structure in Fig.\ref{Schematic}(c). The model Hamiltonian for the K-K$'$ valley excitons is
\begin{equation}
  H = \sum_{i,j} \sum_{\tau,\tau'} t_{ij,\tau\tau'} b_{i\tau}^{\dagger} b_{j\tau'} + \sum_{\{i\}} \sum_{\{\tau\}} U_{i \tau} b_{i \tau_1}^{\dagger} b_{j \tau_2}^{\dagger} b_{i' \tau'_1} b_{j' \tau'_2} \nonumber
\label{EH}
\end{equation}
with $U_{i \tau}\equiv U_{iji'j',\tau_1 \tau_2 \tau'_1 \tau'_2}$ collects the amplitude of both intra-valley and inter-valley scattering of two excitons. $b_{i \tau} = \{B_{i \tau}, D_{i \tau}\}$ can annihilate the bright and dark excitons; $i=A,B$ denotes 1s state of A- and B-excitons and $\tau=\pm 1$ is the valley index. $t_{ij,\tau\tau'}$ accounts for the onsite energy and the electron-hole exchange interaction. 

The exciton dynamics follows the time-evolution, i.e., ${\rm U}(t) = e^{-{\rm i}H (t-t_0)}$. 
The calculations will proceed as usual. 
Inserting into Eq.(\ref{SPPF}) we can obtain the time-domain signal. The case of right-hand circularly polarized s photons \& left-hand circularly polarized i photons is of particular interest hereafter.

At $T=0$, Fig.\ref{QPPF}(a) resolves the biexciton states $|\mathscr{B}_{A,1};\mathscr{B}_{A,-1}\rangle$ at $\bar{v} = 3.180$eV and  $|\mathscr{B}_{B,1};\mathscr{B}_{B,-1}\rangle$ at $\bar{v} = 3.989$eV which thus give the respective binding energies, i.e., $\omega = 2 E - \Delta$ yielding $\Delta_{AA} = 27$meV and $\Delta_{BB} = 16.7$meV. Moreover, one can see additional peaks at $\bar{v} = 3.585$eV. These indicate the A-B biexciton states $|\mathscr{B}_{A,1};\mathscr{B}_{B,-1}\rangle$ and $|\mathscr{B}_{A,-1};\mathscr{B}_{B,1}\rangle$ with a binding energy $\Delta_{AB} = 21.8$meV. The binding energy arises from the dipole-dipole interaction between K- and K$'$-excitons.

With $T > 0$, Fig.\ref{QPPF}(a) shows the exciton dynamics and the conversion between bright and dark excitons. For the resonance at $\bar{v} = 3.180$eV, the increase of its sideband at $\bar{v} = 3.206$eV during the first 100fs shows the conversion between intra-valley bright and dark excitons, i.e., $\mathscr{B}_{A,\pm 1} \rightarrow \mathscr{D}_{A,\pm 1}$. As further extending $T$ to 350fs, the bright exciton transfer emerges obviously thereby, i.e., $\mathscr{B}_{A,1} \rightleftharpoons \mathscr{B}_{A,-1}$ (resulting from the electron-hole exchange) as resolved by the sidebands at $\bar{v} = 3.154$eV, $3.206$eV. One may observe that the sideband intensities are considerably lower than the one at $\bar{v} = 3.206$eV. This is because the peak at $\bar{v} = 3.154$eV arises from the bright exciton transfer only whereas the peak at $\bar{v} = 3.206$eV involves the channels of both bright exciton transfer and bright-dark exciton transfer. Moreover, the bright-dark exciton transfer $\mathscr{B}_{A,\pm 1} \rightarrow \mathscr{D}_{A,\pm 1}$ lasts for about 800fs, as reflected by the continuous increase of the peak $\bar{v} = 3.206$eV intensity until $T=800$fs. 

To highlight the quantum supremacy of using the squeezed light, we essentially make a comparison with the PPF signal using laser pulses [Fig.\ref{QPPF}(b)]. For a pulse duration of 60fs in line with Fig.\ref{QPPF}(a), the spectrum exhibits good temporal but poor spectral resolution, which reveals two broadened features only, at $\bar{v}\approx 3.18,\ 3.59$eV. The spectral lines for bright and dark excitons are thus smeared out, due to the broadband nature of femtosecond lasers. Nevertheless, the signal along the time delay is unable to resolve different exciton transfer channels, and thus does not reflect the real-time dynamics of exctions.

Fig.\ref{QPPF}(a) also depicts the B-exciton dynamics, from the resonances at $\bar{v} = 3.971,\ 3.989,\ 4.007$eV. For instance, the increase of the $\bar{v} = 4.007$eV sideband monitors the conversion between bright and dark excitons, i.e., $\mathscr{B}_{B,1} \rightarrow \mathscr{D}_{B,1}\ \&\  \mathscr{B}_{B,-1} \rightarrow \mathscr{D}_{B,-1}$, during $\sim 800$fs. 
The bright-dark exciton transfer $\mathscr{B}_{B,\pm 1} \rightarrow \mathscr{D}_{B,\pm 1}$ has to be assisted by phonons whereas the one for A excitons, i.e.,  $\mathscr{B}_{A,\pm 1} \rightarrow \mathscr{D}_{A,\pm 1}$ can be spontaneous. Therefore the latter is readily faster, as revealed by the signal along the time delay in Fig.\ref{QPPF}(a).


{\it Conclusions and outlook.}---In summary, we proposed an ultrafast  technique with quantum-correlated photons, for sensing real-time dynamics of materials. Our scheme is presented with a time-energy scale beyond the classical bound---not attainable by laser pump-probe technique---thanks to the quantum squeezing that generates correlated photons. We demonstrated this power through the exciton dynamics in monolayer semiconductor model---the conversion between bright and dark excitons which used to be bottlenecked by the temporal-spectral resolution of the pump-probe scheme. Notably, our results further showed a remarkable enhancement of the signal, e.g., $\sim 6-8$ orders of magnitude greater than the one from entangled photon pairs. Our work provides a proof-of-principle demonstration of unprecedented scales for time-domain spectroscopy to be facilitated with quantum advantage. Our work may open a new frontier for studying the ultrafast dynamics in complex materials.

So far, the time-domain measurement of signals using correlated light reveals a quantum advantage superior to the two-photon absorption, in that the redundant transition pathways solely induced by s/i photons are eliminated. This makes it promising for a selectivity of accessing the material correlation functions. 

\vspace{0.15cm}

\begin{acknowledgments}
We thank Zhe-Yu Ou from City University of Hong Kong, for the fruitful discussions. Z.\ D.\ Z. and J.\ F. gratefully acknowledge the support of the Excellent Young Scientists Fund by National Science
Foundation of China (No. 9240172), the General Fund by National Science Foundation of China (No. 12474364), and the National Science Foundation of China/RGC
Collaborative Research Scheme (No. 9054901). Z.\ D.\ Z. and D.\ L. also acknowledge the financial support from the Guangdong Provincial Quantum Science Strategic Initiative (No. GDZX2205001) and the City University of Hong Kong through the RGMS grant (No. 9229137). S. M. gratefully acknowledges the support of the National Science Foundation through Grant (No. CHE-2246379).
\end{acknowledgments}

\bibliography{sample}

\end{document}